\documentclass[twocolumn,showpacs,preprintnumbers,amsmath,amssymb]{revtex4}
\usepackage{graphicx}% Include figure files
\usepackage{dcolumn}% Align table columns on decimal point
\usepackage{bm}% bold math

\begin{document}

\preprint{APS/123-QED}

\title{Fourier-transfrom Ghost Imaging for pure phase object based on Compressive Sampling algorithm}% Force line breaks with \\

\author{Hui Wang}
\email{ami157@mail.siom.ac.cn}
 %\altaffiliation[Also at ]{Key Laboratory for Quantum Optics and Center for Cold Atom Physics of CAS,\\Shanghai Institution of Optics and Fine Mechanics,\\Chinese Academy of Sciences Shanghai 201800, China.}%Lines break automatically or can be forced with \\

\author{Shensheng Han}%
 \email{sshan@mail.shcnc.ac.cn}
\affiliation{Key Laboratory for Quantum Optics and Center for Cold Atom Physics of CAS,\\Shanghai Institution of Optics and Fine Mechanics,\\Chinese Academy of Sciences Shanghai 201800, China.}

%\affiliation{

\date{\today}% It is always \today, today,

\begin{abstract}
A special algorithm for the Fourier-transform Ghost Imaging (GI) scheme is discussed based on the Compressive Sampling (CS) theory. Though developed mostly in real space, CS algorithm could also be used for the Fourier spectrum reconstruction of pure phase object by setting a proper sensing matrix. This could find its application in diffraction imaging of X-ray, neutron and electron with higher efficiency and resolution. Simulation and experiment results are also presented to prove the feasibility.
\end{abstract}

\maketitle

The diffraction imaging of X-ray, neutron and electron provide significant methods to reveal microstructure \cite{J.Miao, J.Miao2,I. K.,G. J.,B. Reuter,F. Pfeiffer,S. Marchesini}. However, the traditional diffraction imaging by X-ray, neutron and electron in thermal state could only be used in periodic structure imaging. As for the aperiodic structure at the order of the source wavelength, according to the traditional wave theory, the coherent radiation source, such as Free Electron Laser (FEL), is essential for the imaging. However, since neutrons and electrons are fermions, their coherent sources with high brightness are in principle unavailable; although it's possible to obtain a coherent sources of photons with high brightness, the high requirement for brightness could only be met on the Synchrotron Radiation Facilities.

It has been proved that two-photon correlation Ghost Imaging (GI) could realize the diffraction imaging by taking advantages of thermal source \cite{Jing Cheng}, however, it is also limited by the long acquisition time, as the correlation theory calls for mass samples to guarantee the ensemble average. Actually, lots of methods were introduced to improve its Convergence \cite{M. Bache}, however, as long as it still takes the correlation algorithm, the concept of ensemble makes it less effective.

Recently the Compressive Sampling (CS) algorithm attracts more and more attention because of its extraordinary effect of reducing the samples \cite{Justin,E. J.,D. L.,E. J. Candes,J. Romberg,O. Katz}. GI and CS have similarities in extracting imaging: both are prepared with random sensing signal to "express" the imaging and a "point" detector to collect the result of expression. But they also have intrinsic difference: GI is based on accurate "point by point" measure model \cite{Yanhua Shih,M. D¡¯Angelo}, while CS theory has proved that "global random" measure model exhibits higher efficiency in imaging extraction \cite{Justin,E. J.,D. L.,E. J. Candes,J. Romberg,O. Katz}. Therefore, by combing GI and CS, it's possible to develop a brand-new imaging model with higher imaging efficiency and resolution.

At present, CS theory concentrates mostly in the real space, which limits its application in the treatment of grayscale image \cite{Justin,E. J.,D. L.,E. J. Candes,J. Romberg,O. Katz}. While the actual information media waves contain not only amplitude but also phase information, thus the existing CS algorithm could not be used directly to arbitrary GI scheme, especially in the Fourier GI scheme, where most spectrum information relies on the phase part of the field.

In view of these problems, we firstly propose a combination of the Fourier GI and CS in this letter. A recovery algorithm is investigated for the Fourier-Transform Ghost Imaging with Compressive Sampling (GICS) with simulation and experiment results. This algorithm could not only improve the imaging efficiency to reduce the radiation damage of the sample, but may also improve the spectrum resolution compared to the Correlation Ghost Imaging (CGI). Therefore GICS may provide a brand-new diffraction microscopic imaging technic of X-ray, neutron and electron for the aperiodic structure.

According to the CS theory, the algorithm is based on a corresponding relation between the imaging information $X$ and the detect signal $Y$ through a proper sensing matrix $A$ \cite{Justin,E. J.,D. L.,E. J. Candes,J. Romberg}:
\begin{eqnarray}\label{eq1}
 AX=Y
 \end{eqnarray}

If we sample K times, A should be a $K\times N$ matrix, and $Y$ should be a known K-element vector. Then by solving a convex optimization program, it gives an optimal result of $X$ with N-element in the same expressing space as A:
\begin{eqnarray}\label{eq2}
 \min \left\| x \right\|_{l1} {\rm{\    subject\  to\    Y = AX}}
 \end{eqnarray}
Where $x$ are the elements of $X$ in certain express basis, and $\left\| x \right\|_{l_1 }  = \sum\limits_i {\left| {x_i } \right|} $.

\begin{figure}
\includegraphics[width=8.5cm]{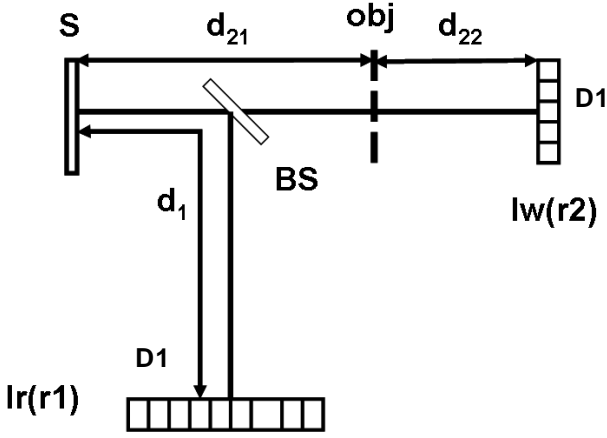}
\caption{The lensless Fourier-transform GI scheme. Beam splitter (BS) is introduced to divide the field into two arms: in the test arm, field propagates freely $d_{21}$ to an object and then $d_{22}$ to array detector $D_2$; in the reference arm, field propagates d1 to another array detector $D_1$.$d_1 = d_{21} + d_{22}$.}
\end{figure}

In our case, the lensless Fourier-transform GI scheme is shown in Fig 1 \cite{Jing Cheng}. The thermal field from the source S is divided by the beam splitter (BS) into test and reference arms. To perform the Fourier-transform GI, there should be $d_1 = d_{21} + d_{22}$ for the scheme.
Based on this scheme, we could express the intensity $I_w(r_2)$ on the test detector D2 as:

\begin{eqnarray}\label{eq3}
 I_w (r_2 ) &\propto& \int\limits_{obj} {dxdx'E^* (x)E(x')} t^* (x)t(x')\\ \nonumber
            & &\times  \exp \left\{ {\frac{{i\pi }}{{\lambda d_{22} }}\left[ {(x - r_2 )^2  - (x' - r_2 )^2 } \right]} \right\}
 \end{eqnarray}

Where the integration is over the object plane. By comparing with the CS theory, the object information is located on the right side of (\ref{eq3}), thus $I_w(r_2)$ could be considered as the known detect signal $Y$. Since our imaging goal is the Fourier-transform of $t(x)$, to process the CS algorithm, it's necessary to establish a corresponding relation like (\ref{eq1}), and the sensing matrix $A$ must be related to the signal $I_r(r_1)$ from the reference detector $D_1$ as the spectrum used to be expected to show up on the array detector $D_1(r_1)$.

If we consider the array of $D_1$ to be large enough, then (\ref{eq3}) could be rewritten approximately as
\begin{widetext}
\begin{eqnarray}\label{eq4}
 I_w (r_2 ) &\propto& \int\limits_{ref} {dr_1 dr_1 'E^* (r_1 )E(r_1 ')} \int\limits_{obj} {dxdx'\exp \left\{ { - \frac{{i\pi }}{{\lambda d_{22} }}\left[ {(x - r_1 )^2  - (x' - r_1 ')^2 } \right]} \right\}}\\ \nonumber
            & & \times t^* (x)t(x')\exp \left\{ {\frac{{i\pi }}{{\lambda d_{22} }}\left[ {(x - r_2 )^2  - (x' - r_2 )^2 } \right]} \right\} \\ \nonumber
            & = & \int\limits_{ref} {dr_1 dr_1 'E^* (r_1 )E(r_1 ')} \exp \left\{ { - \frac{{i\pi }}{{\lambda d_{22} }}\left( {r_1 ^2  - r_1 '^2 } \right)} \right\} \int\limits_{obj} {dx dx'} t^* (x)t(x')\exp \left\{ {\frac{{i2\pi }}{{\lambda d_{22} }}\left[ {(r_1  - r_2 )x - (r_1 ' - r_2 )x'} \right]} \right\} \\ \nonumber
            & = &\int\limits_{ref} {dr_1 dr_1 'E^* (r_1 )E(r_1 ')\exp \left\{ { - \frac{{i\pi }}{{\lambda d_{22} }}\left( {r_1 ^2  - r_1 '^2 } \right)} \right\}} T^* (f_1  = \frac{{r_1  - r_2 }}{{\lambda d_{22} }})T(f_1 ' = \frac{{r_1 ' - r_2 }}{{\lambda d_{22} }}) \\
 \end{eqnarray}
\end{widetext}

Compare (\ref{eq4}) with (\ref{eq1}), the sensing matrix A could then be expressed as $E^* (r_1 )E(r_1 ')\exp \left\{ { - {\raise0.7ex\hbox{${i\pi }$} \!\mathord{\left/ {\vphantom {{i\pi } {\lambda d_{22} }}}\right.\kern-\nulldelimiterspace}\!\lower0.7ex\hbox{${\lambda d_{22} }$}}\left( {r_1 ^2  - r_1 '^2 } \right)} \right\}$.

Worth to note, unlike the sensing matrix $A$ introduced in the bucket real-space GI scheme \cite{O. Katz}, the matrix $A$ here is a $K\times N^2$ one rather than a $K\times N$ one.

This arouses two problems: one is that the CS theory is normally developed in the real space, while the relation (\ref{eq4}) established in the complex space is not directly available in CS algorithm; the other is the unobtainable interference terms of $A$ when $r_1  \ne r_1 '$.

To solve the first problem, $A$ should be redesigned. Define $A=A1+iA2$ and the unknown spectrum $T^* ((r_1  - r_2)/(\lambda d_{22} ))T((r_1 ' - r_2 )/(\lambda d_{22} )) = B_1  + iB_2$. Obviously, $A1$, $A2$ and $B1$, $B2$ are all real matrixes with dimensions of $K\times N^2$ and $N^2\times1$ respectively. Then (\ref{eq4}) could be replaced by two equations:

\begin{eqnarray}\label{eq5}
 A_1 B_1  - A_2 B_2  = I_{\rm{w}}
 \end{eqnarray}

 \begin{eqnarray}\label{eq6}
 A_1 B_2  + A_2 B_1  = 0
 \end{eqnarray}

It seems there're $2N^2$ degree of freedom for X until we realize the symmetry of $A1$ and the anti-symmetry of $A2$. This in return guarantees the symmetry of $B1$ and the anti-symmetry of $B2$ according to the convex optimization (\ref{eq2}) in CS progressing, and make (\ref{eq6}) always hold. Therefore we can rewrite the new sensing matrix $A'$ by combing the elements of $A1$ and $A2$ without introducing extra degree of freedom beyond $N^2$.

 Fig 2 is an example to rebuilt $A'$ for a single sampling: its upper triangular elements are the same as those of $2\times A1$, while the lower triangular ones are the same as $-2\times A1$, and the diagonal line ones are the square root of N intensities $\sqrt {I_r (r_1 )} $ from N points on $D1$. Thus by performing the CS algorithm, as long as we know the exact field distribution on D1, not only the spectrum intensity $\left| {T\left( f \right)} \right|$ is obtainable on the diagonal line, but also its phase information from the non- diagonal elements.

 \begin{figure}
\includegraphics[width=8.5cm]{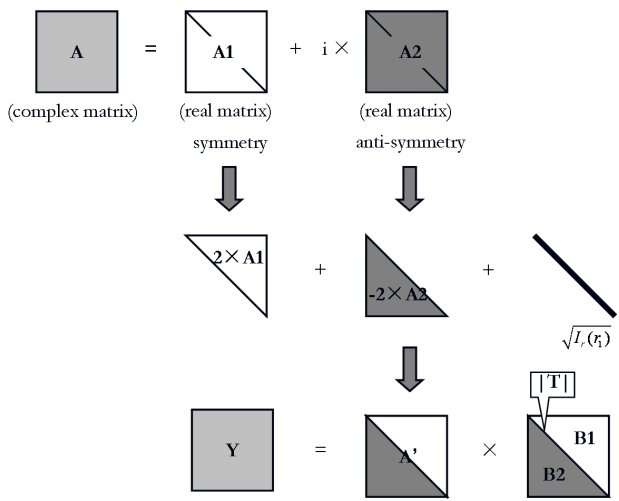}
\caption{An example to rebuilt $A'$.}
\end{figure}

Generally, the only way available to get the exact field distribution on D1 in GI scheme is by calculating for the controllable source scheme or by introducing another known reference field $E'$ for the homodyne detection. However, both are unavailable for the diffraction imaging of X-ray, neutron and electron. Notice the final imaging spectrum result obtained from correlation GI scheme is just $\left| {T\left( f \right)} \right|$, which lies on the diagonal line of the CS solution $X$, corresponding to the known intensity distribution of $D1$ on the diagonal line of $A'$. According to the CS algorithm, the relative distribution of $\left| {T\left( f \right)} \right|$ could be expressed through the known $I_r(r_1)$ of K sampling, while the existence of other non-diagonal elements just make the relation (\ref{eq5}) hold. Therefore, if we only want to get $\left| {T\left( f \right)} \right|$, the accuracy of the non-diagonal elements' values, mainly influencing the convergence, become less important. And it's possible for us to proper conjecture the phase distribution on D1 based on the concrete scheme and translate it into $A'$ followed by normal CS algorithm.

Obviously, (\ref{eq5}) is not a strict equation in the phase conjecture course, just as a non-ideal bucket detector for the real-space imaging \cite{O. Katz}, whose influence to the sensing matrix is ignored. However, this approximation only influences the efficiency to extract the information rather than destroying the expressing basis. Also, there's little relation in applications with strict equation for the CS algorithm to perform, thus it makes little sense to pursue the strict condition for CS algorithm. On the contrary, it's proper to develop the CS algorithm adaptive to relations with different approximation. And still, GICS, as a brand new imaging algorithm instead of CGI, brings improvement of the low efficiency and limit resolution to the traditional CGI.

Our simulation scheme is the same as shown in Fig.1, with source size to be 3 mm, $d_{21}=20 cm, d_{22}=20 cm, d_1=d_{21}+d_{22}$, both $D1$ and $D2$ are array detectors. The object is a one dimension pure phase five-slit, whose transmission function and spectrum intensity $\left| {T\left( f \right)} \right|$ is shown in Fig 3. We simulate the result of GICS with homodyne detection and phase conjecture courses in Fig 4 (a) and (b). For comparison, we also simulate a result of GICS algorithm applied merely to the square root of N intensities from $D1$ (shown in Fig 4 (c)) to show the improved convergency by introducing in the non-diagonal elements. All results are performed through spatial averaging treatment over K=50 sampling \cite{M. Bache}. We could see that GICS could also extract the desired information with higher efficiency even without a strict equation (\ref{eq4}). Besides, the extra non-diagonal elements, though just coming from conjecture, improve the extract efficiency.

\begin{figure}
\includegraphics[width=8.5cm]{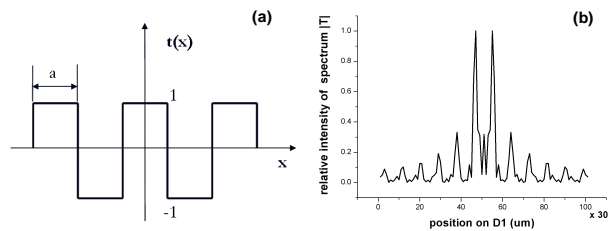}
\caption{Transmission function and spectrum intensity $\left| {T\left( f \right)} \right|$ of the pure phase five-slit object. (a)Transmission function of the object, the slit wide $a$ is 600 $\mu$m.(b)Spectrum intensity $\left| {T\left( f \right)} \right|$ of the object observed from a lens of 30 mm focal length.}
\end{figure}

\begin{figure*}
\includegraphics[width=17cm]{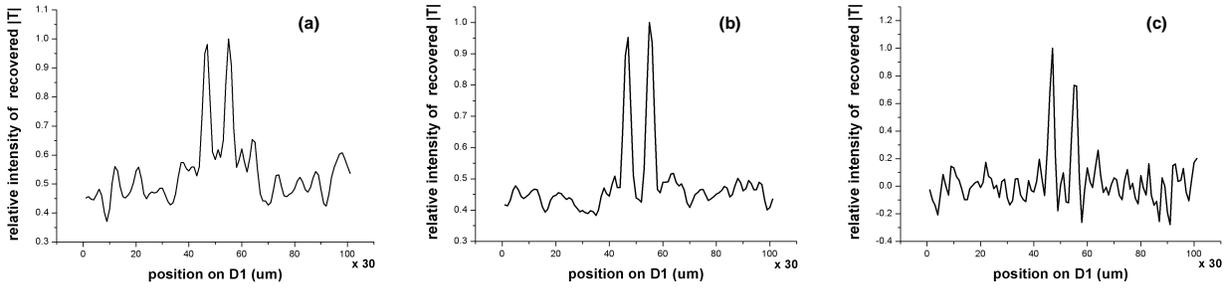}% Here is how to import EPS art%
\caption{\label{fig:wide} Simulation results of GICS with homodyne detection, phase conjecture courses based on the Fig 1 scheme with $d_{21}=,d_{22}=,d_1=d_{21}+d_{22}$. (a)result of GICS with homodyne detection .(b)result of phase conjecture courses.}
\end{figure*}

%\begin{figure}
%\includegraphics[width=8.5cm]{fig3.eps}
%\caption{Simulation result of GICS with homodyne detection and phase conjecture courses based on the Fig 1 scheme with %$d_{21}=,d_{22}=,d_1=d_{21}+d_{22}$. (a)result of GICS with homodyne detection .(b)result of phase conjecture courses.}
%\end{figure}

We also take experiments of the same scheme, where a pseudo-thermal light is produced by permeate a laser beam with diameter of 3 mm through a rotating ground glass, $d_{21}=20 cm,d_{22}=5 cm, d_1=d_{21}+d_{22}$, and the object is a similar one dimension pure phase five-slit with $a = 150 \mu m$. The sampling number K $=$ 100, and the result has been performed through spatial averaging technique.

\begin{figure}
\includegraphics[width=8.5cm]{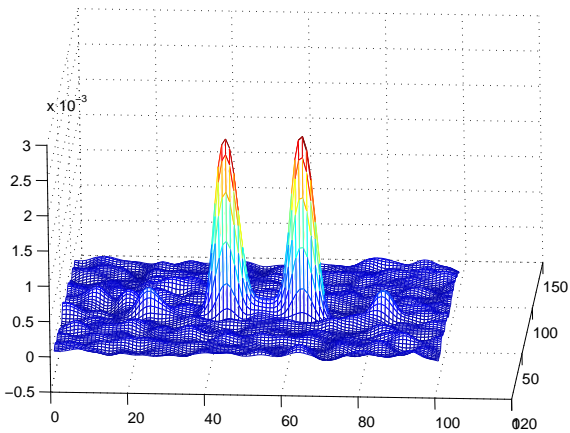}
\caption{Experiment result of GICS based on the GI scheme shown in Fig 1, where $d_{21}=20 cm,d_{22}=5 cm$. The }
\end{figure}

In conclusion, A new algorithm based on the lensless Fourier-transform GICS is presented in this letter. This scheme opens up a new way for the diffraction imaging of X-ray, neutron and electron with higher efficiency and resolution. The GICS, however, is much more than a algorithm revolution. From the point of microscopic essence, quantum fluctuations of field are generally described via high order correlation, which is based on accurate classical "point by point" measurement: the traditional imaging technic is first order correlation (single-photon) system, while CGI is a second order correlation (two-photon) system. CS, on the contrary, stems from a "global random" sensing measurement. By introducing CS algorithm into GI, it also initiates a new discussion on its microscopic quantum mechanism. To find a reasonable explanation for the microscopic essence of GICS is a challenging problem, which will help to develop new imaging model with higher efficiency.

The work is supported by the Hi-Tech Research
and Development Program of China under Grant No.
2006AA12Z115, National Natural Science Foundation of China under Grant Project No.
60877009, and Shanghai Natural Science Foundation under Grant
Project No. 09JC1415000.


\begin{thebibliography}{99}
\bibitem{J.Miao}J.Miao, D.Sayre, and H. N. Chapman, J. Opt.Soc.Am A 15, 1662 (1998).
\bibitem{J.Miao2}J.Miao, P. Charalambous, J. Kirz, and D. Sayre, Nature(London) 400, 342 (1999).
\bibitem{I. K.}I. K. Robinson, I. A. Vartanyants, G. J. Williams, M.A.Pfeifer, and J. A. Pitney, Phys. Rev. Lett. 87, 195505 (2001)
\bibitem{G. J.}G. J. Williams, M. A. Pfeifer, I. A.Vartanyants, and I. K. Robinson, Phys. Rev. Lett. 90,175501 (2003).
\bibitem{B. Reuter}B. Reuter and H. Mahr, Experiments with Fourier transform holograms using 4.49nm x-rays, J. Phys. E., Vol.9(9): 746-751, (1976).
\bibitem{F. Pfeiffer}F. Pfeiffer, T. Weitkamp, O. Bunk and C. David, Nature Physics, 2, 258, (2006).
\bibitem{S. Marchesini}S. Marchesini, H. He, H.N. Chapman, S.P. Hau-Riege, A. Noy, M.R. Howells, U. Weierstall, and J.C.H. Spence, Phys. Rev. B 68, 140101(R) (2003).
\bibitem{Jing Cheng}Jing Cheng, and Shensheng Han. Phys.Rev. Lett. 92, 9 (2004).
\bibitem{M. Bache}M. Bache, E. Brambilla, A. Gatti and L.A. Lugiato, Opt. Express 12, 24(2004)
\bibitem{Justin} Justin Romberg, IEEE SIGNAL PROC MAG. March, 14-20(2008)
\bibitem{E. J.} E. J. Candes and M. B. Wakin, IEEE Sig. Proc. Mag. March, 21-30 (2008).
\bibitem{D. L.}D. L. Donoho and Y. Tsaig, IEEE Trans. Inform. Theory. 54, 4789-4812 (2006).
\bibitem{E. J. Candes}E. J. Candes, J. Romberg, and T. Tao, IEEE Trans. Inform. Theory, 52, 5406-5425 (2006).
\bibitem{J. Romberg}J. Romberg, IEEE Sig. Proc. Mag. March, 14 (2008).
\bibitem{O. Katz}O. Katz, Y. Bromberg, and Y. Silberberg, Appl. phys. Lett. 95, 131110 (2009).
\bibitem{Yanhua Shih}Yanhua Shih, Front. Phys. China, 2(2): 125¨D152, (2007)  
\bibitem{M. D¡¯Angelo}M. D'Angelo and Y.H. Shih, Laser Phys. Lett. 2, 12: 567¨C596 (2005)


\end{thebibliography}
\end{document}